**Excess backgrounds in Dark Matter detectors and physics of glasses**

Author: Sergey Pereverzev

affiliation: LLNL

Abstract

Multiple mechanisms allow energy accumulation in materials and later releases. Interactions between excitations, defects, or other configurations carrying excess energy can lead to avalanche energy releases and more complex phenomena like self-organization and self-replication effects in systems with energy flow. Exact theoretical models of these phenomena are often impossible because of insufficient knowledge of interactions inside materials. Still, comparison and analogies between excess low-energy backgrounds in solid-state detectors, mechanisms of inelastic deformations of crystals, and relaxation processes in glasses allow essential predictions for the background properties and insight into the physics of glasses.
The presence of avalanche-like releases of stored energy is expected for  NaI(Tl). We observed energy accumulation and release processes in the form of delayed random luminescence. We also demonstrated suppression of delayed luminescence by exposure to red light. We cannot yet clearly confirm or refute the presence of small avalanches or excess coincidences in the delayed photon flux. An interplay of fast and delayed luminescence response depends on the type of particles and temperature. One can expect that other environmental factors could cause modulations of low-energy background. More studies of material responses to radiation are required.

**Introduction**
Practically all dark matter and CEvNS detectors demonstrate excess low-energy backgrounds [1]. In this paper, we continue to discuss the role of energy accumulation and release processes in the production of excess backgrounds. Properties of excess backgrounds – a sharp increase in the number of events for lower events energies, an increase of backgrounds with the growth of mechanical stress in solid-state detectors, and an increase of low-energy background with the ionization load in noble-liquid detectors are pointing to the possible common mechanism or scenario for this background production[2]. Interactions between states bearing excess energy can lead to several emerging phenomena. For systems with energy flow, 1977 Noble Laureate  Ilia  Prigogine outlined possibilities of forming dissipative systems, generation of complexity, and transitions from havoc to order [3]. The Self-Organized Criticality theory investigates systems where internal interactions lead to avalanche-like releases of stored energy [4,5]. When relaxational avalanches are present, systems often demonstrate common properties: the spectrum of relaxation events is polynomially decreasing with energy (i.e., catastrophic events are possible), and the power noise spectrum is close to 1/f (pink noise). Reinforcement of relaxation on a small scale (quenching of excitations) can suppress the number of large avalanches[4,5].
Though the properties of materials and excitations and interactions could be different, searching for common scenarios helps understand the effects observed in various types of detectors. The excess low-energy background was observed in practically all low-temperature solid-state experiments looking for dark matter particles and coherent scattering of low-energy neutrinos; several EXCESS workshops were devoted to this matter [1,6]; see also [7]. Yearly modulation of low-energy background in NaI(Tl) scintillator detector was attributed to interaction with dark matter particles by DAMA-LIBRA collaboration [8]. Still, other experiments do not confirm these conclusions. We discuss condensed matter aspects of low-energy background production in solid-state cryogenic detectors and in NaI(Tl) scintillators. The assumption that DAMA-LIBRA modulation includes energy accumulation and releases in the material was first formulated by David Nugren [9].

**Low-energy background in low-temperature solid-state detectors, microcracking and microscopic mechanisms of inelastic deformation of single crystals**

Low-temperature solid-state detectors use different single-crystal sensors: Si, Ge, $SiO_2$, $CaWO_4$, etc., cooled to 10-50 mK temperatures. Experiments look for signatures of nuclear recoils in these crystals using a thermometer attached to the crystal and looking for the temperature spikes, or a superconducting hot phonon sensor/ microbolometer on the surface of the crystal (or array of hot phonon sensors) and looking for the appearance of hot phonon pulses. In parallel, one can look for the emission of photons (scintillation) and electrons/conductivity pulses in the crystal. All types of detectors demonstrate excess low-energy background raising sharply as the energy of the event goes below approximately 10-1 eV[5,6,7] (we will not discuss here questions of recoil energy calibration). In many experiments, the intensity of the low-energy background increases with the increase of thermomechanical stress in the crystal, and additional experiments verify this observation. The other statement is that excess background decreases slowly after the initial cool-down of the detector on a time scale of weeks or months. This relaxation can restart after warming up to several K temperatures- not all the way to room temperature. The above properties of excess backgrounds were attributed to microcracking (see [7] for example) -a process of relief of mechanical stress produced during cool-down by the differences in thermal expansion of materials.

Microscopic mechanisms of small inelastic deformation (flow) of single-crystal samples (Si and Ge most often) were studied under different load conditions, including inhomogeneous loading and micro-indentations [10,11]. It was found that flow has a form of small steps, each consisting of transformation in a microscopic volume of material. These transformations can be changes in crystallographic structure, chemical transformations, the appearance of the twin boundaries, sliding plains, appearance and motion of dislocations. These transformations are dissipative events and should result in heat or hot phonon productions. In many cases, the formation and movement of defects, dislocations, etc., in dielectrics, semiconductors, and metal samples can be accompanied by photon emission and electrons from the sample's surface [12,13]. As we can see, "microcracking" likely has the same meaning as inelastic deformation /flow of crystal material. Thus, the low-energy background we see is a natural relaxation process of mechanical stress release in single crystals and solids in general. As a term, "microcracking" could be misleading, as the production and motion of defects in the crystal are not immediately leading to mechanical breaking or failure. The Inelastic deformation of metals has similar mechanisms, but the limits of plasticity (ductility) can be much more significant for metals.

Below we will discuss other relaxation processes in solids that can generate small discrete energy release events. We argue that the observed excess low-energy backgrounds are a cumulative effect of various glass-like relaxation processes.

**Relaxation processes in glasses and Excess low-energy backgrounds in low-temperature solid-state detectors**

At temperatures below the glass transition, amorphous (disordered) materials are out of mechanical and thermodynamic equilibrium; relaxation processes in the glass state (response to force or other stimuli) became long and dependent on the internal state of the material (i.e., history-dependent). When one applies force for some time and then removes it, slow "back-relaxation" can be present. Such effects mean energy accumulation and releases can be present when the system demonstrates glass-like relaxation properties.

At low temperatures, many subsystems in materials demonstrate complex and history-dependent relaxation properties. Here we provide multiple references on low-temperature physics papers: charges localized on boundaries and interphases in SQUIDs [14], the motion of charges in dielectric substrate probed by single-electron transistor on the surface [15], magnetic moments of impurities in superconductors [16]. The glass-like properties of the relaxation processes become more pronounced with cooling to Ultra-Low Temperatures (below mK); for example, memory effects are present in the dielectric response of glasses at ultra-low temperatures[17]. As temperature became lower, more subtle interactions started to play a role in relaxation dynamics and internal ordering;  nuclear magnetic moments can order inside the superconductor illustrated at [18] as an example

These parallels and analogies between the relaxation process in glasses and processes generating excess background in low-temperature solid-state detectors lead to an important conclusion: any variations of the electric and magnetic fields or irradiation with IR  or microwave radiation can lead to excess energy accumulation in materials at low temperatures. Different non-equilibrium configurations of charges, magnetic moments, nuclear moments, defects, etc., are interacting directly and indirectly (through lattice deformations—i.e., via phonons—or electron systems—i.e., via Ruderman–Kittel–Kasuya–Yosida (RKKY) interactions). So, the author posits avalanche-like relaxation events could also be possible. We will likely start to see these events as we improve the energy sensitivity of our detectors.

**Effect of ionizing radiation on Excess low-energy background**

In addition to mechanical stress, ionizing radiation can deposit excess energy into materials. Examples are thermally stimulated luminescence, electron emission, and conductivity. The energy accumulated in the material due to irradiation leads to photon emission, electron emission from the surface, and generation of current carriers- upon the temperature increase, which leads to thermal activation of stored energy releases. Irradiation produces a different population of energy-bearing states and defects than mechanical stress. Atom displacements can be made not only due to direct momenta transfer from energetic particles but also by producing electronic excitations that initiate chemical reactions between neighboring atoms in the lattice. The formation of molecules leads to vacancies, and the decay of metastable molecules can lead to the formation of vacancy-interstitial atom pairs (Frenkel pairs). Thus, even irradiation with UV light can produce structural defects in crystals.

Thus, we see similarities between the energy deposition in crystal materials by inelastic deformation and ionizing radiation. We also know that many of the produced excitations and defects at low temperatures demonstrate glass-like relaxation properties. The author posits that exposure of the solid-state detector to ionizing radiation or UV  light should lead to delayed low-energy background events with events spectrum and decay time resembling those caused by mechanical stress. Likely, the best way to observe these effects is to expose a solid-state low-temperature detector with low residual mechanical stress to a controllable dose of ionizing radiation at low-background-radiation conditions in an underground experiment.

**Understanding of glasses**

A dominant theoretical approach to glasses based on a tunneling two-level systems model developed back in the 1970s [19,20]. The two-level systems (TLS) model postulates the existence in glasses (in disordered solids) of multi-particle configurations or single particle/electron states with two closely arranged minimal energy configurations or closely spaced energy minima. Tunneling between these configurations/states is responsible for the glass relaxation properties. While TLS models can describe many features of glasses, the TLS model may not be the only explanation suited for describing the observed behaviors [21]. Each TLS is represented with two parameters: energy differences between the two adjacent minima and tunneling probability. The microscopic models for TLS with required parameters are still debated [22]. Two-level systems were assumed to be non-interactive [19,20]; the universality of glass properties in the TLS model originates in the universality of TLS parameter distribution for different materials. On the opposite, interactions between energy-bearing states lead to new emerging phenomena in Prigogine's approach and to the universality of SOC theory. Dug Osheroff [17] also pointed out the importance of internal interactions for understanding ultra-low temperature glasses. We hope that further investigation of excess low-energy backgrounds in the low-energy threshold detectors will help to resolve these apparent contradictions.

**Energy accumulation and release in NaI(Tl)**

David Nygren first suggested that mechanisms responsible for low-energy background in the DAMA-LIBRA experiment can involve effects of energy accumulation and release [9]. The Saint-Gobain company observed that mild UV light exposure of the NaI(Tl) crystal might lead to delayed scintillation pulses (few per second) resembling the effect of irradiation with a few keV energy electrons. These pulses disappear by themselves in several hours or days. This effect can manifest SOC-like dynamics in NaI(Tl), and after-luminescence can be likely suppressed by exposure to red or IR light, as such exposure suppresses thermally stimulated luminescence in alkali-metal halides [2]. At Lawrence Livermore Laboratory, we looked at NaI(Tl) response to UV irradiation under more controllable conditions. We observed a delayed luminescence response lasting several days after exposure to UV light. Exposure to energetic electrons from a $^{60}$Co source leads to delayed luminescence decaying in several hours, see Fig.1. We also find that single energetic particle events and muons are causing delayed emission of photons. The delayed luminescence signals we observed were predominantly random photon emission processes. Saint-Gobain researchers somehow were not pointing out this random flux of uncorrelated single photons. More accurate analysis is required to check for the presence of photon bunching – the appearance of dense emission events above or random coincidences in a random Poisson process. We cannot confirm or refute the Saint-Gobain statement at present. We have tried data acquisition with triggering on coincidences of photon detection by right and left PMTs coupled to the NaI(Tl) crystal and applied a pulse-shape discrimination analysis similar to the study done by the DAMA-LIBRA collaboration. In this case, leakage into the "particle-like" domain is dependent on the left-right coincidence time interval, the choice of the length of the data collecting window after the coincidence, and a choice of "separate event" criteria, i.e., absence of photons for some time intervals before and after the event. Though these filters strongly reduce the number of "candidate particle-like events," we can see some leakage into the "particle domain." A more detailed analysis with an exact replication of DAMA-LIBRA algorithms is required to draw any quantitative conclusions.

The photon flux detected by DAMA-LIBRA and similar experiments consists of delayed luminescence photons (mostly random) and bursts of photon emission produced as a fast/immediate response to external particles. The energy for the delayed photon emission comes from cosmogenic and residual radioactivity, and some energy can come from the residual mechanical stress in the scintillator crystals. The pulse-shape discrimination technique is used to look for the photon-emission bursts caused by nuclear recoils. As one moves to lower-energy recoils, one has fewer photons in the bursts and more chances that a burst-looking "coincidence event" in the random/uncorrelated delayed photon flux can leak through the pulse-shape filters. In addition, because of interactions between states storing excess energy in the material, additional correlation in delayed photon flux could be present that mimics nuclear recoil and pass through our pulse-shape filters (we have not confirmed or refuted this possibility yet). Correlated (avalanche-like) releases of stored energy can be spontaneous, and they also can be activated by small-energy nuclear recoils, like recoils produced by low-energy solar neutrinos. The energy of these recoils is too tiny to produce luminescence photons. But it still exceeds thermal energy, so these recoils can trigger releases of stored energy and produce luminescence. This process is analogous to a suggestion by A. Drukier to use a material containing "micro-explosives" inclusions to detect low-energy nuclear recoils [24].

We have demonstrated (see Fig.1) that exposure to red light can strongly suppress delayed luminescence. It is interesting to check if IR light outcide of PNT sensitivity can be used to suppress delayed luminescence. Then one can look for luminescence under constant illumination of crystals with IR light- to change/suppress backgrounds caused by delayed releases of stored energy.

We know that delayed luminescence depends on the type of primary particles and its decay time depends on the temperature [25,26]. The luminescence response of KI(Tl) can be affected by an applied electric field [27]. This suggests that other environmental factors also can affect delayed luminescence response and cause modulation of the DAMA-LIBRA signal; the time of the maximum intensity of random/uncorrelated delayed luminescence can be different from the maximum of muon flux which pumps energy into the system (this was not yet checked).

Our conclusion is that experimental investigation of the delayed luminescence with the goal of producing a realistic quantitative phenomenological model is required; it is unlikely that we can clarify the mechanism producing modulation of low-energy backgrounds in DAMA-LIBRA experiment without such studies.

**Conclusions.**

When we are looking for rare low-energy interactions of particles with our detectors, we also start to see energy releases by condensed matter and chemical processes, which may resemble nuclear recoils caused by dark matter particles or neutrinos.

We are using elaborate models for dark matter particles and their interactions with detectors, while processes of energy accumulation and releases in materials remain under-investigated.

Investigation of energy accumulation and release effects in detectors is required by using tools and ideas from particle physics and condensed matter physics. This will help to improve exclusion limits for different dark matter particle models and interactions beyond the standard model; for condensed matter physics, such research can help to resolve the unsettled question about the role of interactions of energy-bearing states in glasses and in detector materials and can shed light on the problem of non-thermal noise and decoherence in quantum sensors.

**Acknowledgments**.

This project is supported by the U.S. Department of Energy (DOE) Office of Science/High Energy Physics under Work Proposal Number SCW1508 awarded to Lawrence Livermore National Laboratory (LLNL). LLNL is operated by Lawrence Livermore National Security, LLC, for the DOE, National Nuclear Security Administration (NNSA) under Contract DE-AC52-07NA27344.

**References**

1. EXCESS 2021 workshop: Descriptions of rising low-energy spectra  arXiv:2202.05097 [astro-ph.IM]
2. Sergey Pereverzev, "Detecting low-energy interactions and the effects of energy accumulation in materials", Phys. Rev. D 105, 063002 (2022). https://arxiv.org/abs/2107.14397
3. Ilya Prigogine, "Time, structure and fluctuations", *Science,* Vol. 201, Issue 4358, pp. 777-785, (1978); DOI: 10.1126/science.201.4358.777; or Nobel Lecture, December 8, 1977.
4. Per Back, Cho Tang, and Kurt Weisenfeld, "Self-Organized Criticality: An Explanation of 1/f Noise", Phys. Rev. A, 38, pp. 364-374, (1988).
5. Per Back and Can Chen, "Self-organized criticality", Scientific American, Vol. 264, pp. 46-53, (1991).
6. EXCESS 2022 workshop 15-17 February 2022 EXCESS2022 Workshop (15-17 February 2022) · Indico@KIT (Indico)
7. R. Anthony-Petersen, et.al., "A Stress Induced Source of Phonon Bursts and Quasiparticle Poisoning", arXiv:2208.02790 **[physics.ins-det]**
8. R. Bernabei, et all., "First Model Independent Results from DAMA-LIBRA - Phase 2", Nucl. Phys.Atom.Energy 19 (2018) no.4, 307-325; (arXiv:1805.10486).

9. David Nygren, "A testable conventional hypothesis for the DAMA-LIBRA annual modulation", arXiv:1102.0815 [astro-ph.IM], (2011).
10. Liangchi Zhang, Irena Zarudi, "Towards a deeper understanding of plastic deformation in mono-crystalline silicon", International Journal of Mechanical Sciences 43 (2001).
11. Bradby, J.E., Williams, J.S., Wong-Leung, J. *et al.* Mechanical deformation in silicon by micro-indentation. *Journal of Materials Research* **16,** 15 (2001). https://doi.org/10.1557/JMR.2001.0209
12. W.D. Von Voss and F.R. Brotzen," Electron emission from plastically strained aluminum", J. of Applied Physics, 30,1639 (1959).
13. W.J. Baxter, "A study of plastic deformation by exoelectron emission", Vacuum, V 22, pp.571-575 (1972).
14. S.K. Choi, D.H. Lee, S.G. Louie, J. Clarke: "Localization of metal-induced gap states at the metal-insulator interface: origin of flux noise in SQUIDs and superconducting qubits" Phys. Rev. Lett. 103, 197001 (2009).
15. A. Pourkabirian, M. V. Gustafsson, G. Johansson, J. Clarke, and P. Delsing, "Nonequilibrium Probing of Two-Level Charge Fluctuators Using the Step Response of a Single-Electron Transistor", PRL 113, 256801 (2014).
16. V.M. Galitski, A.I. Larkin, "Spin glass versus superconductivity", Phys. Rev. B 66, 064526 (2002).
17. S. Rogge, D, Natelson, and D. Osheroff, "Evidence for the Importance of Interactions between Active Defects in Glasses," Phys. Rev. Lett. 76, 3136, (1996).
18. S. Rehmann, T. Herrmannsdorfer, F. Pobell, "Interplay of nuclear magnetism and superconductivity in $AuIn_2$", PRL 78, pp.1122-11125 (1997).
19. P. W. Anderson , B. I. Halperin & c. M. Varma "Anomalous low-temperature thermal properties of glasses and spin glasses," Phil. Magazine, 25:1, 1-9 (1972).
20. W. A. Phillips, "*Two-level states in glasses*", Rep. Prog. Phys. 50, 165723 (1987).
21. A.J. Leggett, D.C. Vural, *"Tunneling two-level systems" model of the low-temperature properties of glasses: are "smoking gun" test possible?",* The Journal of Physical Chemistry B, 117, pp. 12966-12971, (2013).
22. C. Muller, J.H. Cole, J. Lisenfeld, *"Towards understanding two-level-systems in amorphous solids: Insights from quantum circuits"*, Reports on Progress in Physics, V.82, 124501 (2019).
23. F. Sutanto, J. Xu,1 S. Pereverzev, and A. Bernstein, "Energy storage and release in sodium iodide scintillation crystals", https://arxiv.org/abs/2208.09974
24. A. K. Drukier, R. L. Fagaly and R. Bielski, "Towards a new class of detectors for dark matter and neutrinos", International Journal of Modern Physics A, Vol. 29, 1443006 (2014)
25. J C Robertson and J G Lynch, "The Luminescent Decay of Various Crystals for Particles of Different Ionization Density", 1961 *Proc. Phys. Soc.* **77** 751
26. C. R. Emighand L. R. Megill, "The Long-Lived Phosphorescent Components of Thallium-Activated Sodiuxn Iodide", Phys. Rev. 93, 1190 (1954).
27. Tetsusuke Hayashi, et.al., "Host-Sensitized Luminescence of KI:TL under an Applied Electric Field", Journal of the Physical Society of Japan, V.33, pp.1018-1023, (1972)

Fig.1

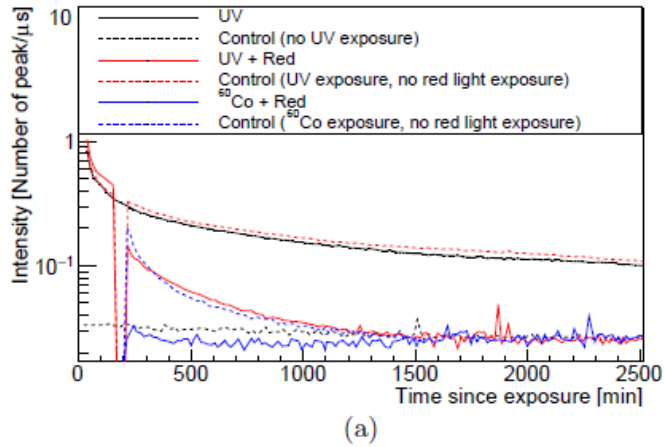

(a)

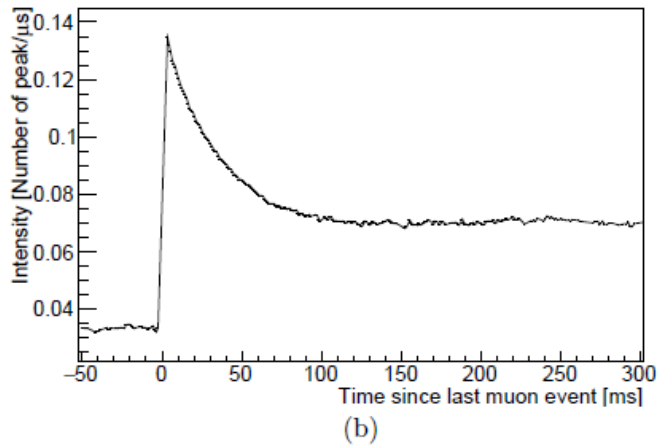

(b)

FIG. 1. The delayed light intensity was monitored over time following UV exposure (a, solid black line) and Co-60 irradiation (a, solid blue line). Three hours after the UV exposure, the crystal was exposed to red light (a, solid red line); and a pronounced decrease in the afterglow intensity was observed. Control runs were performed, in which exposure with the UV or red light sources was omitted, but the same procedure was used for source placement as in the runs with exposure. This check confirmed that the experimental procedure had a negligible contribution to the observed trend (a, dashed lines). We also observed delayed light following large energy depositions in the crystal (b). Note that we omit the data point at the zero time, which is the time of the large ionization event (b).